\documentstyle[12pt]{article}
\newcommand{\alt}{\mathrel{\raisebox{-.6ex}{$\stackrel{\textstyle<}{\sim}$}}}

\def\overlay#1#2{\ifmmode \setbox 0=\hbox {$#1$}\setbox 1=\hbox to\wd 0{\hss
$#2$\hss }\else \setbox 0=\hbox {#1}\setbox 1=\hbox to\wd 0{\hss #2\hss }\fi
#1\hskip -\wd 0\box 1}
\def\case#1/#2{{\textstyle{#1\over#2}}}
\catcode`@=11
\newcount\@tempcntc
\def\@citex[#1]#2{\if@filesw\immediate\write\@auxout{\string\citation{#2}}\fi
  \@tempcnta\z@\@tempcntb\m@ne\def\@citea{}\@cite{\@for\@citeb:=#2\do
    {\@ifundefined
       {b@\@citeb}{\@citeo\@tempcntb\m@ne\@citea\def\@citea{,}{\bf ?}\@warning
       {Citation `\@citeb' on page \thepage \space undefined}}%
    {\setbox\z@\hbox{\global\@tempcntc0\csname b@\@citeb\endcsname\relax}%
     \ifnum\@tempcntc=\z@ \@citeo\@tempcntb\m@ne
       \@citea\def\@citea{,}\hbox{\csname b@\@citeb\endcsname}%
     \else
      \advance\@tempcntb\@ne
      \ifnum\@tempcntb=\@tempcntc
      \else\advance\@tempcntb\m@ne\@citeo
      \@tempcnta\@tempcntc\@tempcntb\@tempcntc\fi\fi}}\@citeo}{#1}}
\def\@citeo{\ifnum\@tempcnta>\@tempcntb\else\@citea\def\@citea{,}%
  \ifnum\@tempcnta=\@tempcntb\the\@tempcnta\else
   {\advance\@tempcnta\@ne\ifnum\@tempcnta=\@tempcntb \else \def\@citea{--}\fi
    \advance\@tempcnta\m@ne\the\@tempcnta\@citea\the\@tempcntb}\fi\fi}
\catcode`@=12

\font\fortssbx=cmssbx10 scaled \magstep2
\hbox to \hsize{
\hskip.5in \raise.1in\hbox{\fortssbx University of Wisconsin - Madison}
\hskip1in$\vcenter{\hbox{\bf MAD/PH/830}
            \hbox{May 1994}}$ }

\vspace{.75in}

\begin{document}
\thispagestyle{empty}

\begin{center}
{\large\bf Singlet quarks beyond the top at the Tevatron?}\\[.4in]
V.~Barger$^a$ and R.J.N.~Phillips$^b$\\[.2in]
\it
$^a$Physics Department, University of Wisconsin, Madison, WI 53706, USA\\
$^b$Rutherford Appleton Laboratory, Chilton, Didcot, Oxon OX11 0QX, UK
\end{center}

\vspace{.5in}

\begin{abstract}
The first evidence for the top quark at the Tevatron may indicate a cross
section higher than the QCD expectation. We consider the possibility that
isosinglet heavy quarks may be contributing to the signal and discuss ways of
testing this possibility.
For example, a charge $\case2/3$ singlet quark, approximately degenerate and
mixing with the top quark, would effectively double the standard top
signals.  A charge $-\case1/3$ singlet quark mixing with the bottom quark
would not affect top signals but would generate excess $Z+{}$multijet
events with a $b$-tag.

\end{abstract}

\newpage

The first evidence for a top-quark signal has just been presented by the CDF
experiment at the Fermilab Tevatron $p\bar p$ collider\cite{cdf}, indicating a
mass $m_t = 174\pm10^{+13}_{-12}$~GeV. An excess of multijet events is found,
containing either two $W$-bosons or a $W$-boson and at least one $b$-jet, where
these $W$-bosons are identified by $W\to\ell\nu$ leptonic decays. The observed
signal rate is somewhat higher than the standard QCD expectations for $p\bar
p\to t\bar t X$ production\cite{ellis}. Although this higher rate may be
attributed to statistical fluctuations or background uncertainties, it has
already encouraged theoretical speculation about possible enhancements of the
$t\bar t$ cross section, such as a color singlet and octet resonances coupled
strongly to top quarks\cite{hill} or a techni-eta resonance\cite{eichten}. A
different new-physics possibility is not that $t\bar t$ production itself is
enhanced but that other heavy quarks are produced and contribute to the
observed signals. A fourth-generation quark (mentioned as a possibility in
Ref.\cite{cdf}) is not particularly attractive, since a fourth light sequential
neutrino is excluded by $Z$ decay data\cite{lep}. However, a theoretically
interesting possibility is the existence of isosinglet quarks that occur for
example in superstring-inspired E$_6$ models\cite{bdpw,rizzo,kane,bw} or other
exotic quarks outside the Standard Model\cite{aguila,ma}. Isosinglet quarks
are among the few classes of new particle that could exist near the electroweak
mass scale without much perturbing the standard analysis of
electroweak radiative corrections. In the present letter we concentrate on the
isosinglet options, the phenomenologies of which have been considered in other
contexts\cite{bdpw,rizzo,kane,bw,aguila,bp}.

In addition to the fermions of the Standard Model (SM), we
address the possibility
that each generation includes either a singlet charge $-{1\over3}$ quark
$Q =x_d, x_s, x_b$ or a singlet charge $\case2/3$ quark  $Q=x_u, x_c, x_t$.
These new singlet quarks are color-triplets and are produced by standard
QCD subprocesses; their production rates are exactly those for SM
quarks of the same masses. They decay via mixing with SM quarks of the
same charge into $qW$, $q'Z$ and $q'H$ channels, where $q(q')$ is a lighter
quark and $H$ is the SM Higgs boson; if the mixing is small the decay
interactions and branching fractions are simply
related \cite{bdpw,rizzo,kane,bw,aguila}:
\begin{equation}
    B(Q \to qW) : B(Q \to q'Z) : B(Q \to q'H) \simeq 2 : 1 : 1 \;,
\end{equation}
apart from kinematic factors that are $\simeq 1$ for $m_Q >> M_W,M_Z,m_H$.
We assume for simplicity that the mixing occurs mostly within the same
generation, in
which case $q(q')$ is the corresponding light quark: $x_d \to uW, dZ, dH$,
etc. The Higgs considerations generalize somewhat beyond the SM; in the
minimal supersymmetric extension\cite{hhg}, for example, if there is only
one light Higgs boson (the charged and other neutral Higgses
comparatively heavy) then its couplings are close to those of the SM.
For present purposes we shall assume $m_H \alt M_Z$; if $H$ is very much
heavier than this, it will be suppressed in decays of singlet quarks
near the top mass.

   Since these new quarks introduce new decay modes, it may be pertinent now
to mention some further aspects of the CDF top search\cite{cdf}. Compared
to standard expectations and the measured top production rate, CDF
reports a deficit of a few events in the $W+4\,$jets background rate
(that might be explained by fewer top events) and an excess of
2 events in the tagged $Z+4\,$jets channel compared with $0.64\pm0.06$
expected.  Both effects could be statistical fluctuations\cite{cdf}.

   Consider first the case of charge $-\case1/3$ (``down-type") singlet quarks
and suppose that at least one of them has mass near $m_t$ and is
pair produced at the Tevatron at a rate comparable with $t\bar t$.
Its decay branching fractions\cite{bdpw,rizzo,kane,bw,aguila} are then
approximately
\begin{equation}
\begin{array}{lcccc}
B(x_d \to uW, dZ, dH) & \simeq &  {1\over2}, & {1\over4}, & {1\over4} \;, \\
B(x_s \to cW, sZ, sH) & \simeq &  {1\over2}, & {1\over4}, & {1\over4} \;, \\
B(x_b \to tW, bZ, bH) & \simeq &      0,     & {1\over2}, & {1\over2}  \;,
\end{array}
\end{equation}
with $x_b\to tW$ forbidden by kinematics.
It follows immediately that only the $x_d$ and $x_s$ options yield large
$qW$ decay fractions.  These additional singlet $Q\to qW$ contributions
are superficially similar to $t \to bW$ decays, but differ in important ways.\\
(i) The untagged single-$W$ signal rate is only about half of that for
$t\bar t$ production with the same mass (where $bW$ decays are 100 $\%$),
assuming that the $qZ$ and $qH$ hadronic decays look passably like hadronic
$qW$.\\
(ii) The untagged two-$W$ signal rate is only about a quarter of that
for $t\bar t$.\\
(iii) The final quark $q = u$ or $c$ is not a $b$-quark; however $c$-quarks
can sometimes satisfy the lepton or vertex criteria used in $b$-tagging
and therefore masquerade as $b$.\\
(iv) The charged lepton in the subsequent $W \to \ell \nu$ decay has a
different kinematical distribution relative to the initial and final quark
momenta\cite{bp} (to be precise, it corresponds to the neutrino
distribution in $t \to bW \to b\ell\nu$); however, in small data samples
this distribution cannot be accurately determined.

There is therefore some potential for $x_s \to cW$ decays to mimic b-tagged
top signals, but at rates reduced by the $x_s \to cW$ branching fraction
and by the tag-factor for $c$-jets.  Thus the tendency would be  to
increase the untagged $W+4\,$jets background much more than the apparent
top signal; but in the CDF data this background seems already too low,
so to this extent the $Q=x_s$ hypothesis is disfavored.
On the other hand, the equally populated $x_s \to sZ, sH$ modes give rise
to $x_s\bar x_s \to s\bar s WZ(WH,ZZ,HH,ZH)$ final states, with no
counterparts in top decays.  As noted above, $WZ$ and $WH$ can contribute
to the single-$W$ top signal, since $Z\to q\bar q$ or $H\to b\bar b$ dijet
decays can mimic $W\to q\bar q'$.    But in cases where $Z$ is identified by
$Z\to\ell\bar\ell$ or $Z\to\nu\bar\nu$ (missing $p_T$), excess $csWZ$ events
with two extra hard quark jets could be seen.  Also the $s\bar sZZ$ and $s\bar
sZH$ modes contribute excess $Z+{}$multijets events with high $b$-tag
probability (from $Z,H\to b\bar b$);  the CDF excess of tagged $Z+4\,$jet
events might be explained in this way.  However, each $(Z\to \ell\bar\ell)jjjj$
event implies approximately six $(W\to \ell\nu)jjjj$ events from other
$x_s\bar x_s$ decay modes, so explaining the $Z+4\,$jet excess in this way
would make the $W+4\,$jets deficit more acute.

   Alternatively, if we address the $Z+4\,$jets excess alone, the case
$Q=x_b$ becomes attractive.  It generates no top-like signal nor
unwanted $W+{}$multijets background, but gives new $b\bar bZH$, $b\bar bZZ$ and
$b\bar bHH$ final states, of which the first two could easily provide tagged
$Z+4\,$jets events (and incidentally a possible Higgs
signal\cite{kane,bw,aguila}).

   Consider next the case of charge $\case2/3$ (``up-type") singlet quarks
and suppose that at least one of them has mass near $m_t$ and is
pair produced at the Tevatron at a rate comparable with $t\bar t$.
Its decay branching fractions\cite{aguila} are then approximately
\begin{equation}
\begin{array}{lcccc}
B(x_u \to dW, uZ, uH) & \simeq & {1\over2} , & {1\over4} , & {1\over4} \;,\\
B(x_c \to sW, cZ, cH) & \simeq & {1\over2} , & {1\over4} , & {1\over4} \;,\\
B(x_t \to bW, tZ, tH) & \simeq & 1 , & 0  , & 0  \;
\end{array}
\end{equation}
with $x_t\to tZ,\,tH$ forbidden by kinematics.
Each of these options yields large $qW$ decay fractions, but the capacity
to mimic top decays depends on the particular case.\\
(i) $x_u\bar x_u$ and $x_c\bar x_c$ production give untagged single-$W$
and two-$W$ signals similar to $t\bar t$, with reduced rates (like
down-type singlets) but similar lepton distribution to $t$ decay (unlike
down-type singlets).  However, the associated quarks are $d$ and $s$,
so these signals would only get into a b-tagged sample via mis-tagging.
There would be $b$-tagged $Z+4\,$jet contributions but
the $W+{}$multijets background deficit would
get worse.  There is little to recommend these cases.\\
(ii) $x_t\bar x_t$ production however gives signals almost identical to
$t\bar t$.  The $x_t$ decays to $Z$ and $H$ are suppressed (but could proceed
at some level via mixing with other generations). A major difference
between $x_t\bar x_t$ and standard  $t\bar t$
signals is in the lifetime\cite{life}: the $t$ decays before hadronization
can happen, so effects like spin depolarization and quarkonia formation are
suppressed for $t\bar t$; $x_t$ lives much longer, due to the small $t-x_t$
mixing, so such effects are allowed for $x_t\bar x_t$ states.\\
(iii) $x_t$-onia are an interesting subject in themselves.
Quarkonium states can be produced via gluon fusion at hadron colliders.
Their single-quark decay modes would be suppressed by the small $x_t-t$
mixing and hence various annihilation decays, such as
$ZZ$, $Z\gamma$, $ZH$, $HH$ and $H\gamma$ \cite{onium},
might be detectable.

To summarize, the central question is whether singlet quarks $Q$ with mass
near $m_t$ may be contributing a significant part of the CDF top signals.
We conclude as follows.\\
(a) The cases $Q=x_d,x_b,x_u,x_c$ cannot contribute significantly to the
CDF top signals; their single-$W$ and two-$W$ signals are reduced by the
$Q\to qW$ branching fraction (that vanishes for $x_b$) and further
suppressed by $b$-tagging.\\
(b) The case $Q=x_s$ is less suppressed by $b$-tagging and can contribute
a small fraction of the top signal.  E$_6$ models can accommodate such
charge $-\case1/3$ singlets.\\
(c) The case $Q=x_b$ is interesting for a different reason; it
contributes nothing to the CDF top signals nor $W+{}$multijets backgrounds,
but it can provide tagged $Z+4\,$jets events as seen by CDF, most of which
would be $b\bar bZH$ events containing a Higgs signal\cite{kane,bw,aguila}.
E$_6$ models can accommodate this case too.\\
(d) This $Z+\,$multijet production could be important in other contexts,
e.g. as an extra source of events with high missing transverse energy
$\overlay/E_T$, that might be confused with supersymmetry signals. Two
events with high $\overlay/E_T$ were reported in early CDF
data\cite{ptmiss}.\\
(e) The case $Q=x_t$ can almost exactly duplicate the top signals;
for mass $m_{x_t}=m_t$ it would double the top signal rate.  However,
we know of no popular models containing this case.\\
(f) We recall that all heavy singlet scenarios imply heavy quarkonium
possibilities\cite{onium}.\\
(g) Event ratios in the more interesting cases may be summarized
approximately:
\begin{equation}
\begin{array}{lrcr}
x_s\bar x_s \Rightarrow & ccWW : csWZ : csWH:ssZH:ssZZ:ssHH
 &\simeq& 4:4:4:2:1:1 \\
x_b\bar x_b \Rightarrow & bbZH:bbZZ:bbHH
 & \simeq&      8:4:4 \\
x_t\bar x_t \Rightarrow & \multicolumn{1}{l}{bbWW }
 & \simeq &      \multicolumn{1}{l}{16}
\end{array}
\end{equation}
(h) In all these down-type and up-type singlet scenarios, it is understood
that the combined $t\bar t$ plus $Q\bar Q$ events would not simply be
distributed like a standard top signal alone.  Beside the questions of
lepton momentum and decay width mentioned above, the presence of two
(generally different) masses would broaden many distributions such as the
reconstructed top mass and the apparent $t\bar t$ invariant mass.\\

\begin{flushleft}{\bf Acknowledgments}\end{flushleft}
This research was supported in part by the U.S.~Department of Energy under
Contract No.~DE-AC02-76ER00881 and in part by the University of Wisconsin
Research Committee with funds granted by the Wisconsin Alumni Research
Foundation.


\begin{thebibliography}{99}
\frenchspacing

\bibitem{cdf} CDF collaboration: F.~Abe et al, FERMILAB-PUB-94/097-E; the
D0 experiment also presented a top candidate event in
S.~Abachi et al., Phys. Rev. Lett. {\bf 72}, 2138 (1994).


\bibitem{ellis} R.K.~Ellis, Phys.Lett. {\bf B259}, 492 (1991); E.~Laenen,
J.~Smith and W.L.~van Neerven, ibid. {\bf B321}, 254 (1994).

\bibitem{hill} C.~Hill and S.~Parke, FERMILAB-Pub-93/397-T.

\bibitem{eichten} E.~Eichten and K.~Lane, FERMILAB-Pub-94/007-T.

\bibitem{lep} The LEP collaborations: ALEPH, DELPHI, L3, OPAL, Phys. Lett.
{\bf B276}, 247 (1992); SLD collaboration, Mod. Phys. Lett. {\bf A8},
2237 (1993).

\bibitem{bdpw} V.~Barger, N.G.~Deshpande, R.J.N.~Phillips and K.~Whisnant,
Phys. Rev. {\bf D33}, 1912 (1986).

\bibitem{rizzo} T.G.~Rizzo, Phys. Rev. {\bf D34}, 1438 ( 1986); J.L.~Hewett and
T.G.~Rizzo, Z.~Phys. {\bf C34}, 49 (1987); Phys. Rep. {\bf 183}, 193 (1989).

\bibitem{kane}F.~del Aguila, E.~Laermann and P.~Zerwas, Nucl. Phys. {\bf B297},
1 (1988); F.~del~Aguila, G.L.~Kane and M.~Quiros, Phys. Rev. Lett. {\bf 63},
942 (1989).

\bibitem{bw}V.~Barger and K.~Whisnant, Phys. Rev. {\bf D41}, 2120 (1990).

\bibitem{aguila} F.~del Aguila, L.~Ametller, G.L.~Kane and J.~Vidal, Nucl.
Phys. {\bf B334}, 1 (1990).

\bibitem{ma} V.~Barger and E.~Ma, University of Wisconsin-Madison report
MAD/PH/811 (1993).

\bibitem{bp} V.~Barger and R.J.N.~Phillips, Phys. Rev. {\bf D41}, 52 (1990).

\bibitem{hhg} For a review see J.F.~Gunion, H.E.~Haber, G.L.~Kane and
S.W.~Dawson, ``The Higgs Hunter's Guide", Addison-Wesley, Reading, MA, 1990.

\bibitem{life} V.~Barger, H.~Baer, K.~Hagiwara and R.J.N.~Phillips, Phys.
Rev. {\bf D30}, 947 (1984); I.~Bigi, Y.~Dokshitzer, V.~Khose, J.~K\"uhn and
P.~Zerwas, Phys. Lett. {\bf B181}, 157 (1986); V.~Barger, J.~Ohnemus and
R.J.N.~Phillips, Int. J. Mod. Phys. {\bf A4}, 617 (1989).

\bibitem{onium} V. Barger et al., Phys Rev. Lett. {\bf 57}, 1672 (1986); Phys.
Rev. {\bf D35}, 3366 (1987); J.L.~Hewett and T.G.~Rizzo, Phys. Rev. {\bf D35},
2194 (1987).

\bibitem{ptmiss} CDF collaboration: F.~Abe et al., Phys. Rev. Lett.
{\bf 69}, 3439 (1992).


\end{thebibliography}
\end{document}